\documentstyle[aps,preprint,epsfig]{revtex}
\tightenlines

\begin{document}

\title{
The predictability problem in systems
with an uncertainty in the evolution law}

\author{G. Boffetta$^1$, A. Celani$^1$, M. Cencini$^2$, G. Lacorata$^3$
and A. Vulpiani$^2$}
\address{$^1$ Dipartimento di Fisica Generale, Universit\`a di Torino,
Via Pietro Giuria 1, 10125 Torino, Italy \\
Istituto Nazionale Fisica della Materia, Unit\`a di Torino Universit\`a \\
Istituto di Cosmogeofisica del CNR, C. Fiume 4, 10133 Torino, Italy}
\address{$^2$Dipartimento di Fisica, Universit\`a di  Roma
"la Sapienza", \\
Piazzale Aldo Moro 5, 00185 Roma, Italy \\
Istituto Nazionale Fisica della Materia, Unit\`a di Roma 1}
\address{$^3$Dipartimento di Fisica, Universit\`a dell' Aquila, \\
Via Vetoio 1, 67010 Coppito, L'Aquila, Italy \\
Istituto di Fisica dell'Atmosfera, CNR \\
Via Fosso del Cavaliere, 00133 Roma, Italy}

\date{\today}
\maketitle

\begin{abstract}
The problem of error growth due to the incomplete knowledge 
of the evolution law
which rules the dynamics of a given physical system is addressed.
Major interest is devoted to the analysis of error amplification
in systems with many characteristic times and scales.
The importance of a proper parameterization of fast scales in
systems with many strongly interacting degrees of freedom is
highlighted and its consequences for the modelization
of geophysical systems are discussed.
\end{abstract}


\section{Introduction}
\label{sec:1}
The ability to predict the future state of a system, given its present state,
stands at the foundations of scientific knowledge  with relevant implications
from an applicative point of view in geophysical and astronomical sciences.
In the prediction of the evolution of a system, e.g. the atmosphere,
we are severely limited by the fact that we do not know
with arbitrary accuracy the evolution equations and the
initial conditions of the system.
Indeed, one integrates a mathematical model given by
a finite number of equations. The initial condition,
a point in the phase space of the model, is determined only
with a finite resolution (i.e. by a finite number of observations)
(Monin 1973).

Using concepts of dynamical systems theory, there have been some
progresses in understanding the growth of an uncertainty during
the time evolution.
An infinitesimal initial uncertainty
($\delta_{0} \rightarrow 0$) in the limit of long times
($t \rightarrow \infty$) grows exponentially in time with a typical rate
given by the leading Lyapunov exponent $\lambda$,
$|\delta x(t)| \sim \delta_0 \exp(\lambda t)$.
Therefore if our purpose is to forecast the system within a tolerance
$\Delta$, the future state of the system can be predicted
only up to the {\it predictability time}, given by:
\begin{equation}
T_{p} \sim \frac{1}{\lambda} \ln \left(
\frac{\Delta}{\delta_{0}}\right) \, .
\label{eq:1.1}
\end{equation}
In literature, the problem of predictability with respect to
uncertainty on the initial conditions is referred to as {\it
predictability of the first kind}.

In addition, in  real systems we must also cope with the lack
of knowledge of the evolution equations.
Let us consider a system described by a differential equation:
\begin{equation}
{d \over d t} \, {\bf x}(t)={\bf f}({\bf x},t) \, , \; \; \; \;
{\bf x, f} \in {\cal R}^{n}  \, .
\label{eq:1.2}
\end{equation}
As a matter of fact we do not know exactly the equations, and
we have to devise a model which is different from the true dynamics:
\begin{equation}
{d \over dt} \, {\bf x}(t)={\bf f}_{\epsilon}({\bf x},t)
\; \; \; \; \mbox{where} \; \; \;
{\bf f}_{\epsilon}({\bf x},t)={\bf f}({\bf x},t)+\epsilon
\delta{\bf f}({\bf x},t) \, .
\label{eq:1.3}
\end{equation}
Therefore, it is  natural to wonder about the relation between the
true evolution ({\em reference} or {\em true} trajectory ${\bf x}_{T}(t)$)
given by
(\ref{eq:1.2}) and  that one effectively computed ({\em perturbed} or
{\em model} trajectory ${\bf x}_{M}(t)$) given by (\ref{eq:1.3}).
This problem is referred to as {\it predictability of the second kind}.

Let us make some general remarks. At the foundation of the second kind
predictability problem there is the issue of {\it structural
stability} (Guckenheimer et al. 1983): since the evolution laws are
known only with finite precision it is highly desirable that at least
certain properties are not too sensitive to the details of the
equations of motion.  For example, in a system with a strange
attractor, small generic changes in the evolution laws should not
change drastically the dynamics (see Appendix \ref{app:0} for a simple
example with non generic perturbation).

In chaotic systems the effects of a small generic uncertainty on the
evolution law are similar to those due to the finite precision on the
initial condition (Crisanti et al. 1989). The model trajectory of the
perturbed dynamics diverges exponentially from the reference one with
a mean rate given by the Lyapunov exponent of the original system.
The statistical properties (such as correlation functions and temporal
averages) are not strongly modified.  This last feature has been
frequently related to the {\it shadowing lemma}
(Guckenheimer et al. 1983; Ott 1993):
almost all trajectories of the true system can be
approximated by a trajectory of the perturbed system starting
from a slightly different initial condition.
However, as far as we know, the shadowing lemma can be proven
only in special cases and therefore it cannot be
straightforwardly invoked to explain the statistical reproducibility 
in a generic case.
In addition, in real systems the size of an uncertainty on the evolution
equations is determinable only {\it a posteriori}, based on the
ability of the model equations to reproduce some of the features of the
phenomenon.

In dynamical systems theory, the problems
of first and second kind predictability
is essentially understood in the limit of infinitesimal
perturbations.
However even in this limit we must also
consider the fluctuations of the rate of expansion
which can lead to relevant
modifications of the predictability time (\ref{eq:1.1}),
in particular for strongly intermittent systems
(Benzi et al. 1985; Paladin et al. 1987; Crisanti et al. 1993).

As far as finite perturbations are considered, the leading Lyapunov
exponent is not relevant for the predictability issue.  In presence of
many characteristic times and spatial scales the Lyapunov exponent is
related to the growth of small scale perturbations which saturates on
short times and has very little relevance for the growth of large
scale perturbations (Leith and Kraichnan 1972; Monin 1973; Lorenz
1996).  To overcome this shortcoming, a suitable
characterization of the growth of non infinitesimal perturbations, in
terms of the Finite Size Lyapunov Exponent (FSLE),
has been recently introduced (Aurell et al. 1996 and
1997).

Also in the case of second kind predictability one has often to deal
with errors which are far from being infinitesimal.  Typical examples
are systems described by partial differential equations
(e.g. turbulence, atmospheric flows). The study of these systems is
performed by using a numerical model with unavoidable severe
approximations, the most relevant of which is the necessity to cut
some degrees of freedom off; basically, the small scale variables.

The aim of this paper is to analyze the effects of limited resolution
 on the large scale features.
This raises two problems:
in first place one has to deal with perturbations of the evolution 
equations which in general cannot be considered small;
second, the parameterization of the
unresolved modes can be a subtle point.
We shall show that the Finite Size Lyapunov Exponent is able to
characterize the effects of uncertainty on the evolution laws.
Moreover we shall discuss the typical difficulties arising
in the parameterization of the
unresolved scales.

This paper is organized as follows. In section \ref{sec:2} we report some
known results about the predictability problem of the second kind and
recall the definition of the FSLE.
In section \ref{sec:3} we present numerical results on simple
models. In section \ref{sec:4} we consider more complex systems with
many characteristic times.
Section \ref{sec:5} is devoted to summarize the results.
In Appendix \ref{app:0} we illustrate a simple example of 
structural unstable system.
In Appendix \ref{app:1} we describe the method for the computation
of the FSLE and in Appendix \ref{app:2} we discuss the problem of the
parameterization of the unresolved variables.

\section{EFFECTS OF A SMALL UNCERTAINTY ON THE EVOLUTION LAW }
\label{sec:2}
In the second kind predictability problem, we can distinguish three 
general cases depending on the original dynamics.
In particular, equation (\ref{eq:1.2}) may display:

(i) trivial attractors:
asymptotically stable fixed points or attracting periodic
orbits;

(ii) marginally stable fixed points or periodic/quasi-periodic orbits
 as in integrable Hamiltonian systems;

(iii) chaotic behavior.

In case (i) small changes in the equations of motion do not modify
the qualitative features of the dynamics.  Case (ii) is not generic
and the outcome strongly depends on the specific perturbation $\delta {\bf f}$,
i.e. it is not structurally stable.  In the chaotic case (iii) one
expects that the perturbed dynamics is still chaotic. 
In this paper we will consider only this latter case.

Let us also mention that, in numerical computations of evolution equations
(e.g. differential equations), there are two unavoidable sources of errors:
the finite precision representation of the numbers
which causes the  computer phase space to be necessarily discrete and 
the round-off which introduces a sort of noise.
Because of the discrete nature of the phase space of the system
studied on computer, orbits numerically computed have to be periodic.  
Nevertheless the period is usually very large, apart for 
very low computer precision (Crisanti et. al 1989).
We do not consider here this source of difficulties. The round-off
produces on eq. (\ref{eq:1.2}) a perturbation which can be written as
$\delta{\bf f}({\bf x},t) = {\bf w}({\bf x})\, {\bf f}({\bf
x},t)$ and $\epsilon \sim 10^{-\alpha}$ ($\alpha=$number of digits
in floating point representation) where ${\bf w}=O(1)$ is an unknown
function which may depend on ${\bf f}$ and on the software of the
computer (Knut 1969).  In general, the round-off error is very small and may have,
as much as the noise, a positive role, as underlined by Ruelle (1979),
in selecting the physical probability measure, the so-called {\it
natural measure}, from the set of ergodic invariant measures.

In chaotic systems the effects of a small uncertainty on the evolution law
is, for many aspects, similar to those due to imperfect knowledge of initial
conditions.  This can be understood by the following example.
Consider the Lorenz equations (Lorenz 1963)
\begin{equation}
\begin{array}{lll}
\displaystyle{ \frac{dx}{dt} }&=& \sigma(y-x) \nonumber \vspace{5pt}\\
\displaystyle{ \frac{dy}{dt} }&=& Rx-y-xz \nonumber \vspace{5pt}\\
\displaystyle{ \frac{dz}{dt} }&=&xy-bz . \vspace{5pt}
\end{array}
\label{eq:2.1}
\end{equation}
In order to mimic an experimental error in the determination of the
evolution law we consider a small error $\epsilon$ on the parameter $R$:
$R \rightarrow R+\epsilon$. Let us consider the difference
$\Delta {\bf x}(t)={\bf x}_{M}(t)-{\bf x}_{T}(t)$ with, for simplicity,
$\Delta {\bf x}(0)=0$, i.e. we assume a perfect knowledge of the
initial conditions.
One has, with obvious notation:
\begin{equation}
{d \Delta {\bf x} \over d t}={\bf f}_{\epsilon}({\bf x}_{M})-
{\bf f}({\bf x}_{T}) \simeq
\frac{\partial {\bf f}}{\partial {\bf x}} \, \Delta {\bf x}
+\frac{\partial {\bf f}_{\epsilon}}{\partial R} \, \epsilon \, .
\label{eq:2.2}
\end{equation}
At time  $t=0$ one has $|\Delta {\bf x}(0)|=0$, therefore 
$|\Delta {\bf x}(t)|$ grows initially 
only by the effect of the second term in (\ref{eq:2.2}).
At later times, when  $|\Delta {\bf x}(t)|\approx O(\epsilon)$ 
the leading term of (\ref{eq:2.2}) becomes the first one, 
and we recover the first kind predictability problem
for an initial uncertainty $\delta_0 \sim \epsilon$.
Therefore, apart from an initial (not particularly interesting) growth,
which depends strongly on the specific perturbation,
the evolution of $<\log(|\Delta {\bf x}(t)|)>$ follows the usual
linear growth with the slope given by the leading Lyapunov exponent.
Typically the value of the Lyapunov exponent computed by using
the model dynamics differs from the true one by a small amount
of order $\epsilon$, i.e. $\lambda_{M}=\lambda_{T}+O(\epsilon)$
(Crisanti et al. 1989).

This consideration applies only to infinitesimal perturbations. 
The generalization to finite perturbations 
requires the extension of the Lyapunov exponent to finite errors.
Let us now introduce the Finite Size Lyapunov Exponent
for the predictability of finite perturbations.
The definition of FSLE $\lambda(\delta)$ is given in terms of the ``doubling
time'' $T_{r}(\delta)$, that is the time a perturbation of initial
size $\delta$ takes to grow by a factor $r$ ($>1$):
\begin{equation}
\lambda(\delta)=\left<
  \frac{1}{T_{r}(\delta)}\right >_{t} \ln \,r
\label{eq:2.4}
\end{equation}
where $\langle\cdots\rangle_{t}$ denotes average with respect to the natural
measure, i.e. along the trajectory
(see Appendix \ref{app:1}).
For chaotic systems, in the limit of infinitesimal perturbations
($\delta \rightarrow 0$) $\lambda(\delta)$ is nothing but
the leading Lyapunov exponent $\lambda$ (Benettin et al. 1980).
Let us note that the above definition of $\lambda(\delta)$
is not appropriate to discriminate cases with $\lambda=0$ and $\lambda<0$,
since the predictability time is positive by definition.
Nevertheless this is not a limitation as long as we deal with
chaotic systems.

In many realistic situations the error growth for infinitesimal
perturbations is dominated by the fastest scales, which are typically
the smallest ones (e.g. small scale turbulence).  When $\delta$ is no
longer infinitesimal, $\lambda(\delta)$ is given by the fully
nonlinear evolution of the perturbation. In general $\lambda(\delta)
\leq \lambda$, according to the intuitive picture that large
scales are more predictable. Outside the range of scales in which the
error $\delta$ can be considered infinitesimal, the function
$\lambda(\delta)$ depends on the details of the dynamics and in
principle on the norm used.  In fully developed turbulence one has the
universal law $\lambda(\delta) \sim \delta^{-2}$ in the inertial range
(Aurell et al. 1996 and 1997).  It is remarkable that this prediction,
which can be obtained within the multifractal model for turbulence, is
not affected by intermittency and it gives the law originally proposed
by Lorenz (1969).
The behavior of $\lambda(\delta)$ as a function of $\delta$ gives
important information on the characteristic times and scales
of  the system and it has been also applied
to passive transport in closed basins
(Artale et al. 1997).

Let us now return to the example (\ref{eq:2.1}).
We compute $\lambda_{TT}(\delta)$, the FSLE for the true equations,
and $\lambda_{TM}(\delta)$, the FSLE computed following the 
distance between
one true trajectory and one model trajectory starting at the same point.
These are shown in Figure \ref{fig1}.
The true FSLE $\lambda_{TT}(\delta)$ displays a plateau
indicating a chaotic dynamics with leading Lyapunov exponent
$\lambda \simeq 1$.
Concerning the second kind predictability, 
for $\delta > \epsilon$ the second term in (\ref{eq:2.2}) becomes
negligible and we observe the transition to the Lyapunov exponent
$\lambda_{TM}(\delta) \simeq \lambda_{TT}(\delta) \simeq \lambda$.
In this range of errors the model system recovers the intrinsic
predictability of the true system. For very small errors, 
$\delta < \epsilon$, where the growth of the error is dominated 
by the second term in (\ref{eq:2.2}), we have
$\lambda_{TM}(\delta) > \lambda_{TT}(\delta)$.

This example shows that it is possible to recover
the intrinsic predictability of a chaotic system even 
in presence of some uncertainty in the model equations.

The relevance of the above example is however limited by the fact that
(\ref{eq:2.1}) does not involve different scales. To investigate
the effect of spatial resolution on predictability let us consider
the advection of Lagrangian tracers in a given Eulerian field. We study
a time-dependent, two dimensional, velocity field given by the
superposition of large scale (resolved) eddies and small scale
(possibly unresolved) eddies.

The streamfunction we consider is a slight modification of a model
originally proposed for chaotic advection in Rayleigh-B\'enard
convection (Solomon et al. 1988):
\begin{equation}
\Psi(x,y,t)=\psi(x,y,t;k_L,\omega_L, B_L)+
\epsilon \cdot \psi(x,y,t;k_S,\omega_S, B_S)
\label{eq:2.6}
\end{equation}
with
\begin{equation}
\psi(x,y,t;k,\omega, B)={1 \over k} \sin \left\{ k \left[
x+B\, \sin(\omega t)
\right] \right\} \sin\left(k y\right)
\label{eq:2.7}
\end{equation}

The first term represents the large-scale flow, i.e. the resolved
part of the flow, the second one mimics
 the unresolved small scale term and
$\epsilon$ measures the relative amplitude.
We choose
$k_S \gg k_L$ and $\omega_S \gg \omega_L$ in order to have a sharp separation
of space and time scales.

The Lagrangian tracers evolve according to the equations:
\begin{equation}
{dx \over dt}= - {\partial \psi \over \partial y}\,, \;\;\;
{dy \over dt}={\partial \psi \over \partial x}\,.
\label{eq:2.8}
\end{equation}
We use the complete stream function (\ref{eq:2.6}) for the true
dynamics and only the large-scale term
for the model dynamics.

The time dependence induces chaotic motion and diffusion in the $x$
direction, without inserting any noise term (Solomon et al. 1988).
For what concerns $\lambda_{TT}$ one observes three regimes (Figure
\ref{fig2}).  For very small errors $\delta<2\pi/k_S$ the exponential
separation is ruled by the fastest scale and $\lambda_{TT} \simeq
\lambda$, i.e. we recover the Lyapunov exponent of the system.  At
intermediate errors we observe a second small plateau 
corresponding to the large-scale term for the model dynamics.
For larger errors, $\delta>2\pi/k_L$ one has 
$\lambda_{TT} \sim \delta^{-2}$, i.e. diffusive behavior
(see Artale et. al 1997).

The model FSLE $\lambda_{TM}(\delta)$ cannot recover the small scale
features: for $\delta \ll \epsilon$ we observe the scaling
$\lambda_{TM}(\delta) \sim \delta^{-1}$ which can be understood by the
following argument. In this region the distance between the reference
and the true trajectories grows as $d \delta/d t \sim \epsilon$ and
thus, by a dimensional estimate, one has:
\begin{equation}
\lambda_{TM}(\delta) \sim {1 \over T_r(\delta)} \sim
{\epsilon \over \delta} \, .
\label{eq:2.5}
\end{equation}
Nevertheless, for larger $\delta$ the model fairly captures the small
plateau displayed by $\lambda_{TT}(\delta)$ which corresponds to the
slow time scale, then at large $\delta$ (i.e. for $\delta$ greater
than the $\delta>2\pi/k_L$) we recover the diffusive behavior with
the correct diffusion coefficient.  This last feature can be
understood by the fact that the diffusion coefficient, being an
asymptotic quantity of the flow, is not influenced by the details
of small scale structures.

This example is rather simple: large scales do not interact with 
the small ones and the number of degrees of freedom is very small.
Therefore in this case the crude elimination of the small scale component
does not prevent the possibility of a fair description of 
large scale features. 

In the following we will consider more complex situations, in which
strongly interacting degrees of freedom with different characteristic 
times are involved.
In these cases the correct parameterization of the unresolved
modes is crucial for the prediction of large scale behavior.

\section{Systems with two time scales}
\label{sec:3}

Before analyzing in detail the effects of non infinitesimal
perturbations of the evolution laws in some specific models let us
clarify our aims.  We consider a dynamical system written
in the following form:
\begin{equation}
\begin{array}{lll}
\displaystyle{ \frac{d{\bf x}}{dt} }&=& {\bf f}({\bf x},{\bf y}) \vspace{5pt}\\
\displaystyle{ \frac{d{\bf y}}{dt} }&=& {\bf g}({\bf x},{\bf y})\,,
\label{eq:3.1}
\end{array}
\end{equation}
where ${\bf f}, {\bf x}\in {\cal R}^{n}$ and
${\bf g},{\bf y}\in {\cal R}^{m}$, in general $n \neq m$.
Now, let us suppose that the fast variables ${\bf y}$ 
cannot be resolved:
a typical example are the subgrid modes in PDE discretizations.
In this framework, a natural question is:
how must we parameterize the unresolved modes (${\bf y}$) in order to
predict the resolved modes~(${\bf x}$)~?

As discussed by Lorenz
(1996), to reproduce -- at a qualitative level -- a given phenomenology, 
e.g. the ENSO phenomenon, one can drop out the small scale features
without negative consequences.  But one unavoidably fails in
forecasting the ENSO (i.e. the actual trajectory) without taking into 
account in a suitable way the small scale contributions.

An example in which it is relatively simple to develop a model for
the fast modes is represented by skew systems:
\begin{equation}
\begin{array}{lll}
\displaystyle{ \frac{d{\bf x}}{dt} }&=& {\bf f}({\bf x},{\bf y}) \vspace{5pt}\\
\displaystyle{ \frac{d{\bf y}}{dt} }&=& {\bf g}({\bf y})
\end{array}
\label{eq:3.2}
\end{equation}
In this case, the fast modes $({\bf y})$ do not depend on the
slow ones $({\bf x})$.
One can expect that in this case, neglecting the fast variables
or parameterizing them with a suitable stochastic process,
should not drastically affect the prediction of the slow variables
(Boffetta et al. 1996).

On the other hand, if ${\bf y}$ feels some feedback from
${\bf x}$, we cannot simply neglect the unresolved modes.
In Appendix \ref{app:2} we discuss this point in detail.
In practice one has to construct an effective 
equation for the resolved variables:
\begin{equation}
{d {\bf x} \over d t}= {\bf f}_{M}({\bf x},{\bf y({\bf x})}) \,,
\end{equation}
where the functional form of ${\bf y({\bf x})}$ and ${\bf f}_{M}$
are found by phenomenological arguments and/or by numerical studies
of the full dynamics.

Let us now investigate an example with a recently introduced toy 
model of the atmosphere circulation
(Lorenz 1996; Lorenz et al. 1998)
including large scales $x_{k}$ (synoptic scales)
and small scales $y_{j,k}$ (convective scales):
\begin{equation}
\begin{array}{lll}
\displaystyle{ {d x_{k} \over dt} }& = & - x_{k-1} \left(x_{k-2}-x_{k+1}\right)
- \nu x_{k} + F - \sum_{j=1}^{J} y_{j,k} \vspace{5pt}\\
\displaystyle{ {d y_{j,k} \over dt} } & = & - c b y_{j+1,k}
\left(y_{j+2,k}-y_{j-1,k}\right) - c \nu y_{j,k} + x_{k} \\
\end{array}
\label{eq:3.6}
\end{equation}
where $k=1,...,K$ and $j=1,...,J$. As in
(Lorenz 1996)
we assume periodic boundary conditions on $k$ ($x_{K+k}=x_{k}$,
$y_{j,K+k}=y_{j,k}$) while for $j$ we impose
$y_{J+j,k}=y_{j,k+1}$. The variables $x_k$ represent
some large scale atmospheric quantities in $K$ sectors extending on 
a latitude circle, while the $y_{j,k}$ represent 
quantities on smaller scales in $J \cdot K$ sectors.
The parameter $c$ is the ratio between
fast and slow characteristic times and $b$ measures the relative amplitude.

As pointed out by Lorenz, this model
shares some basic properties with more realistic models of the atmosphere.
In particular, the non-linear terms, which model the advection, are
quadratic and conserve the total kinetic energy
$\sum_k (x_k^2 + \sum_j y_{j,k}^2)$ in the unforced ($F=0$),
inviscid ($\nu=0$) limit; the linear terms containing $\nu$ mimic dissipation 
and the constant term $F$ acts as an external forcing  preventing
the total energy from decaying.

If one is interested in forecasting the large scale behavior of the 
atmosphere by using only the slow variables, a natural choice for the
model equations is: 
\begin{equation}
{d x_{k} \over dt} = - x_{k-1} \left(x_{k-2}-x_{k+1}\right)
- \nu x_{k} + F - G_{k}({\bf x})\,,
\label{eq:3.7}
\end{equation}
where $G_{k}({\bf x})$ represents the parameterization of the fast
components in (\ref{eq:3.6})
(see Appendix \ref{app:2}).

The FSLE for the true system
(Boffetta et al. 1998)
is shown in Figure \ref{fig3} and displays the two characteristic
plateau corresponding to fast component (for $\delta \ll 0.1$) and
slow component for large $\delta$ .
Figure \ref{fig3} also shows
what happens when one simply neglects the fast components $y_{j,k}$
(i.e. ${\bf G}({\bf x})=0$).
At very small $\delta$ one has
$\lambda_{TM}(\delta) \simeq \delta^{-1}$ as previously discussed.
For large errors we observe that, with this rough approximation,
we are not able to capture the characteristic predictability of
the original system.
More refined parameterizations in terms of stochastic processes
with the correct probability distribution function and
correlation times do not improve the forecasting ability.

The reason for this failure 
is due to the presence of a feedback term in the equations (\ref{eq:3.6})
which induces strong correlations between
the variable $x_{k}$ and the unresolved coupling $\sum_{j=1}^{J} y_{j,k}$.
For a proper parameterization of the unresolved variables we
follow the strategy discussed in Appendix \ref{app:2}. Basically we adopt
\begin{equation}
G({\bf x}) = \nu_e x_k\,,
\label{eq:3.8}
\end{equation}
in which $\nu_e$ is a numerically
determined parameter.
Figure \ref{fig3} shows that, although small scale are not resolved,
the large scale predictability is well reproduced and one has
$\lambda_{TM}(\delta) \simeq \lambda_{TT}(\delta)$ for large $\delta$.
We conclude this section by observing that the proposed parameterization
(\ref{eq:3.8}) is a sort of eddy viscosity parameterization.

\section{Large scale predictability in a turbulence model}
\label{sec:4}
We now consider a more complex system which mimics the energy cascade
in fully developed turbulence.
The model is in the class of the so called {\it shell models}
introduced some years ago for a dynamical
description of small-scale turbulence. For a recent review on shell
models see
Bohr et al. 1998.
This model has relatively few degrees of freedom but involves 
many characteristic scales and times.
The velocity field is assumed isotropic and it is decomposed
on a finite set of complex velocity components $u_{n}$ representing
the typical turbulent velocity fluctuation on a ``shell'' of
scales $\ell_{n}=1/k_n$.
In order to reach very high Reynolds number with a moderate number
of degrees of freedom, the scales are geometrically spaced as
$k_n=k_0 2^n$  ($n=1,...N$).

The specific model here considered has the form
(L'vov et al. 1998)
\begin{equation}
{d u_n \over d t} = i \left(k_{n+1} u^*_{n+1} u_{n+2} -
{1 \over 2} k_{n} u^*_{n-1} u_{n+1} +
{1 \over 2} k_{n-1} u_{n-2} u_{n-1} \right) -
\nu k_{n}^{2} u_{n} + f_{n}
\label{eq:4.1}
\end{equation}
where $\nu$ represent the kinematic viscosity and $f_n$ is a forcing
term which is restricted only to the first two shells (in order to 
mimic large scale energy injection).

Without entering in the details, we recall that the Shell Model (\ref{eq:4.1})
displays an energy cascade {\em \'a la} Kolmogorov
from large scales (small $n$) to dissipative
scales ($n \sim N$) with a statistical stationary energy flux.
Scaling laws for
the average velocity components are observed:
\begin{equation}
\langle | u_n^{p} | \rangle \simeq k_n^{-\zeta_p}
\label{eq:4.2}
\end{equation}
with exponents close to the Kolmogorov 1941 values $\zeta_p=p/3$.

From a dynamical point of view, model (\ref{eq:4.1}) displays
complex chaotic behavior which is responsible of the small deviation
of the scaling exponents (intermittency) with respect to the
Kolmogorov values.
Neglecting this (small) intermittency effects, a dimensional estimate
of the characteristic time (eddy turnover time) for scale $n$ gives
\begin{equation}
\tau_n \simeq {\ell_n \over |u_n|} \simeq k_n^{-2/3} \, .
\label{eq:4.3}
\end{equation}

The scaling behavior holds up to the Kolmogorov scale
$\eta=1/k_d$ defined as the scale at which the dissipative
term in (\ref{eq:4.1}) becomes relevant. The Lyapunov exponent
of the turbulence model can be estimated as the fastest characteristic 
time $\tau_d$ and one has the prediction
(Ruelle 1979)
\begin{equation}
\lambda \sim {1 \over \tau_d} \sim Re^{1/2}
\label{eq:4.4}
\end{equation}
where we have introduced the Reynolds number $Re \propto 1/\nu$.
It is possible to predict the behavior of the FSLE by observing that 
the faster scale $k_n$ at which an error of size $\delta$ is still
active (i.e. below the saturation) is such that $u_n \simeq \delta$.
Thus $\lambda(\delta) \sim 1/\tau_n$ and, using Kolmogorov scaling,
one obtains
(Aurell et al. 1996 and 1997)
\begin{equation}
\lambda_{TT}(\delta) \sim \left\{
\begin{array}{lcl}
 \lambda \; \; & \hbox{for}& \; \delta \leq u_d \\
 \delta^{-2}   \; \; & \hbox{for}& \; u_d \leq \delta \leq u_0
\end{array}\right.
\label{eq:4.5}
\end{equation}
To be more precise there is an intermediate range between the
two showed in (\ref{eq:4.5}). For a discussion on this point see
(Aurell et al. 1996 and 1997).

In order to simulate a finite resolution in the model, we consider
a modelization of (\ref{eq:4.1}) in terms of an eddy viscosity
(Benzi et al. 1998)
\begin{equation}
{d u_n \over d t} = i \left(k_{n+1} u^*_{n+1} u_{n+2} -
{1 \over 2} k_{n} u^*_{n-1} u_{n+1} +
{1 \over 2} k_{n-1} u_{n-2} u_{n-1} \right) -
\nu^{(e)}_{n} k_{n}^{2} u_{n} + f_{n}
\label{eq:4.6}
\end{equation}
where now $n=1,...,N_M<N$ and the eddy viscosity, restricted to the
last two shells, has the form
\begin{equation}
\nu^{(e)}_{n} = \kappa {|u_n| \over k_n} \left(\delta_{n,N_{M}-1}+
\delta_{n,N_M}\right)
\label{eq:4.7}
\end{equation}
where $\kappa$ is a constant of order $1$ (see Appendix \ref{app:2}).
The model equations (\ref{eq:4.6}) are the analogous of large eddy
simulation (LES) in Shell Model which is one of the most popular
numerical method for integrating large scale flows. Thus, although
Shell Models are not realistic models for large scale geophysical
flows (being nevertheless a good model for small scale turbulent fluctuations),
the study of the effect of truncation in term of eddy viscosity is
of general interest.

In Figure~\ref{fig4.1} we show $\lambda_{MM}(\delta)$, i.e.
the FSLE computed for the model equations (\ref{eq:4.6})
with $N=24$ at different resolutions
$N_{M}=9,15,20$.
A plateau is detected for small amplitudes of the error $\delta$,
corresponding to the leading Lyapunov exponent, which
increases with increasing resolution -- being proportional to
the fastest timescale -- according to
$\lambda \sim k_{N_M}^{2/3}$.
At larger $\delta$ the curves collapse onto the $\lambda_{TT}(\delta)$,
showing that large-scale statistics of the model is not affected
by the small-scales resolution.

The capability of the model to predict satisfactorily
the statistical features of the ``true'' dynamics is not anyway
determined by $\lambda_{MM}(\delta)$ but by
$\lambda_{TM}(\delta)$, which is shown in Figure~\ref{fig4.2}.

Increasing the resolution $N_M=9,15,20$ towards the fully
resolved case $N=24$ the model improves,
in agreement with the expectation that $\lambda_{TM}$ approaches $\lambda_{TT}$
for a perfect model.
At large $\delta$ the curves practically coincide, showing that
the predictability time for large error sizes (associated with large scales)
is independent on the details of small-scale modeling.
Better resolved models achieve  $\lambda_{TM} \simeq \lambda_{TT}$
for smaller values of the error $\delta$.

\section{Conclusions}
\label{sec:5}

In this Paper the effects of the uncertainty of the evolution laws 
on the predictability properties are investigated and quantitatively
characterized by means of the Finite Size Lyapunov Exponent.
In particular, we have considered systems involving several 
characteristic scales and times. In these cases, it is rather 
natural to investigate what is the effect of small scale parameterization
on large scale dynamics.

It has been shown that in systems where there is  a negligible feedback on
the small scales by the large ones, the dynamics of the former ones can be
thoroughly discarded, without affecting the statistical features of
large scales and the ability to forecast them. 
On the other side, when this feedback is present, the
crude approximation of cutting the small scale variables off is
no longer acceptable.  In this case one has to model the action
of fast modes (small scales) on slow modes (large scales) with some
effective term, in order to recover a satisfactory forecasting of large
scales.  The renowned eddy-viscosity modelization is an instance of
the general modeling scheme that has been here discussed.

\section{Acknowledgments}
We thank L. Biferale for useful suggestions and discussions.  This
work was partially supported by INFM (Progetto Ricerca Avanzata TURBO)
and by MURST (program 9702265437). A special acknowledgment goes to
B. Marani for warm and continuous support.

\appendix
\section{An example of structural unstable system}
\label{app:0}
In order to see that a non generic perturbation, although very ``small'',
can produce dramatic changes in the dynamics, let us discuss a simple example
following (Berkooz 1994; Holmes et al. 1996).
We consider the one-dimensional chaotic
map $x_{t+1}=f(x_{t})$ with $f(x)=4 x$ mod $1$, 
and a perturbed version of it:
\begin{equation}
f_p(x)=\left\{ 
\begin{array} {ll}

8x-\frac{9}{2} & \;\;\; x \in \left[\frac{5}{8},\frac{247}{384}\right]  \\ 
\\
\frac{1}{2}x+\frac{1}{3} & \;\;\;x \in \left[\frac{247}{384},
\frac{265}{384}\right]\\ 
\\
8x-\frac{29}{6} &  \;\;\; x \in \left[\frac{265}{384},\frac{17}{24}\right] \\
\\
4x\;{\mbox {mod}}\;1 & \;\;\;{\mbox{otherwise}}\,.
\end{array}
\right.    
\label{eq:ap0}
\end{equation}
The perturbed map is identical to the original outside the interval
$[5/8,17/24]$, and the perturbation is very small in $L_{2}$ norm.
Nevertheless, the fixed point $x=\frac{2}{3}$, which is unstable in the
original dynamics, becomes stable in the perturbed one. Moreover it is a
{\it global attractor} for $f_p(x)$, i.e. almost every point in
$[0,1]$ asymptotically approaches $x=\frac{2}{3}$ (see Figure~\ref{fig0}).

Now, if one compares the trajectories obtained iterating $f(x)$ or $f_p(x)$ 
it is not difficult to understand that orbits starting 
outside $[5/8,17/24]$ remain identical
for a certain time but unavoidably they differ 
utterly in the long time behavior.
It is easy to realize that the transient chaotic behavior of 
the perturbed orbits can be rendered arbitrarily long by reducing the 
interval in which the two dynamics differ.
This example shows how even an ostensibly small perturbation (in 
usual norms) can modify dramatically the dynamics.

\section{Computation of the Finite size Lyapunov exponent}
\label{app:1}
In this appendix we discuss in detail the computation of the Finite
Size Lyapunov Exponent for both continuous dynamics (differential
equations) and discrete dynamics (maps).

The practical method for computing the FSLE goes as follows.
Defined a given norm for the distance $\delta(t)$ between the
reference and perturbed trajectories, one has to define a series of thresholds
$\delta_{n}=r^{n}\delta_{0}$ ($n=1,\dots, N$),
and to measure the ``doubling times'' $T_{r}(\delta_{n})$ that a
perturbation of
size $\delta_{n}$ takes to grow up to $\delta_{n+1}$.
The threshold rate $r$ should not be taken too large, because
otherwise the error has to grow through different scales before
reaching the next threshold. On the other hand, $r$ cannot be
too close to one, because otherwise the doubling time would be
of the order of the time step in the integration. In our examples
we typically use $r=2$ or $r=\sqrt 2$. For simplicity $T_{r}$
is called ``doubling time'' even if $r \neq 2$.

The doubling times $T_{r}(\delta_{n})$ are obtained by following
the evolution of the
separation from its initial size $\delta_{min} \ll \delta_0$ up to
the largest threshold $\delta_{N}$.
This is done by
integrating the two trajectories of the system starting at an initial
distance $\delta_{min}$. In general, one must choose
$\delta_{min} \ll \delta_{0}$, in order to allow the direction of the
initial perturbation to align with the most unstable direction in the
phase-space. Moreover, one must pay attention to keep
$\delta_{N} < \delta_{saturation}$, so that all the thresholds
can be attained ($\delta_{saturation}$ is the typical distance
of two uncorrelated trajectory, i.e. the size of the attractor).
For the second kind predictability problem, i.e. the computation
of $\lambda_{TM}(\delta)$, one can safely take $\delta_{min}=0$ because
this do not prevent the separation of trajectories.

The evolution of the error from the initial value $\delta_{min}$ to
the largest threshold $\delta_{N}$ carries out a single error-doubling
experiment. At this point one rescales the model trajectory at the
initial distance $\delta_{min}$ with respect to the true trajectory
and starts another experiment.
After ${\cal N}$ error-doubling experiments, we can estimate the
expectation value of some quantity $A$ as:
\begin{equation}
\langle A \rangle_{e} = {1 \over {\cal N}} \sum_{i=1}^{\cal N} \, A_i \, .
\label{eq:ap1}
\end{equation}
This is not the same as taking the time average as in (\ref{eq:2.4})
because different error doubling experiments may takes different times.
Indeed we have
\begin{equation}
\langle A \rangle_{t} = {1 \over T} \int_0^T \, A(t) dt =
{\sum_i A_i \tau_i \over \sum_i \tau_i} =
{\langle A \tau \rangle_{e} \over \langle \tau \rangle_{e}} \, .
\label{eq:ap2}
\end{equation}
In the particular case in which $A$ is the doubling time itself we have
from (\ref{eq:2.4}) and (\ref{eq:ap2})
\begin{equation}
\lambda(\delta_n) = {1 \over \langle T_{r}(\delta_n) \rangle_{e}} \ln r \, .
\label{eq:ap3}
\end{equation}

The method described above assumes that the distance between the two
trajectories is continuous in time. This is not true for maps of for
discrete sampling in time and the method has to be slightly modified.
In this case $T_{r}(\delta_n)$ is defined as the minimum time at which
$\delta(T_r) \ge r \delta_n$. Because now $\delta(T_r)$ is a fluctuating
quantity, from (\ref{eq:ap2}) we have
\begin{equation}
\lambda(\delta_n) = {1 \over \langle T_{r}(\delta_n) \rangle_{e}}
\left\langle \ln \left( {\delta(T_r) \over \delta_n} \right) \right\rangle_{e} \
, .
\label{eq:ap4}
\end{equation}

We conclude by observing that the computation of the FSLE is not
more expensive than the computation of the Lyapunov exponent by
standard algorithm. One has simply to integrate two copies of the
system (or two different systems for second kind predictability)
and this can be done also for very complex simulations.

\section{Parameterization of small scales}
\label{app:2}
Typically a realistic problem (e.g. turbulence) involves many interacting
degrees of freedom with different characteristic times.
Let us indicate with ${\bf z}$ the state of the system under consideration,
with an evolution law:
\begin{equation}
\frac{{\rm d}{\bf z}}{{\rm d} t}={\bf F}({\bf z}) \,,\:\: {\bf F},{\bf z} \in
{\cal R}^{N}\,.
\label{eq:ap2.1}
\end{equation}
The dynamical variables ${\bf z}$ can be split in two sets:
\begin{equation}
{\bf z}=({\bf x},{\bf y})\,,
\label{eq:ap2.2}
\end{equation}
where ${\bf x} \in {\cal R}^{n}$ and ${\bf y} \in {\cal R}^{m}$
($N=n+m$), being respectively ${\bf x}$ and ${\bf y}$
the ``slow'' and ``fast'' variables.
The distinction between slow and fast variables is often largely arbitrary.

The evolution equation (\ref{eq:ap2.1}) is divided into two blocks,
the first one containing the dynamics of  the slow variables,
the second one associated with the dynamics of the fast variables

\begin{equation}
\left\{  \begin{array} {c}
\displaystyle{\frac{d{\bf x}}{dt} } = {\bf F_{1}}({\bf x})+
{\bf F_{2}}({\bf x},{\bf y})  \vspace{5pt} \\

\displaystyle{\frac{d{\bf y}}{dt} } = {\bf \tilde{F}_{1}}({\bf x},{\bf y})+
{\bf \tilde{F}_{2}}({\bf y})
\end{array}
\right.
\label{eq:ap2.3}
\end{equation}

If one is interested only in the slow variables it is necessary to write an
``effective'' equation for ${\bf x}$.
As far as we know there is only one case for which it is simple to find 
 the effective equations for ${\bf x}$.
If the characteristic times of the fast variables are much smaller than
those ones of the ${\bf x}$ (adiabatic limit), one can write:
\begin{equation}
{\bf y}=<{\bf y}>+{\mbox {\boldmath $\eta$}}(t)
\label{eq:ap2.4}
\end{equation}
where {\boldmath $\eta$} is a Wiener process, i.e.
a zero mean Gaussian process with
\begin{equation}
< \eta_{i}(t)\eta_{j}(t^{'})>=<\delta y_{i}^2> \delta_{ij} \delta(t-t^{'})\,.
\label{eq:ap2.5}
\end{equation}
Therefore one obtains for the slow variables:
\begin{equation}
\frac{{\rm d}{\bf x}}{{\rm d} t} = {\bf F_{1}}({\bf x})+
\delta {\bf F_{1}}({\bf x})+\delta {\bf W}({\bf x},{\mbox {\boldmath $\eta$}})
\label{eq:ap2.6}
\end{equation}
where $\delta {\bf F_{1}}({\bf x})={\bf F_{2}}({\bf x},<\!\!\!{\bf y}\!\!\!>)
+\delta{\bf F_{2}}$,
$\delta F_{2,j}=1/2 \sum_i {\partial}^2 F_{2,j}/\partial y_{j}^2
<\delta y_{i}^2>$ and
$\delta W_i=\sum_i \left. \partial F_{2,j}/\partial{y_j}
\right|_{<{\bf y}>} \eta_i(t)$.
Basically the slow variables ${\bf x}$ obey to a non linear Langevin
equation. 

Here the role of the fast degrees of freedom becomes relatively
simple: they give small changes to the drift ${\bf F_{1}}\rightarrow
{\bf F_{1}}+ \delta{\bf F_{1}}$ and a noise term
$\delta {\bf W}({\bf x},{\mbox {\boldmath $\eta$}})$.
We remark that the validity of the above argument is rather limited.
Even if one has a large time scale separation,  the statistics of the
fast variables can be very far from the Gaussian distribution. In
particular, in system with feedback (${\bf \tilde{F}}_{1} \ne 0$) one
cannot model the fast variable ${\bf y}$ independently of the resolved
${\bf x}$.

In the generic situation the construction of the effective equation
for ${\bf x}$ requires to follow phenomenological arguments which
depend on the physical mechanism of the particular problem.  For
example, for the Lorenz '96 model discussed in sect. \ref{sec:3}, 
where $F_{2,k}({\bf x},{\bf y})=\sum_{j=1,J} y_{j,k}$, we
use the following procedure for the parameterization of the fast
variables and the building of the effective eq. for ${\bf x}$.
Instead of assuming (\ref{eq:ap2.4}) we mimic the fast variables in
terms of the slow ones:
\begin{equation}
{\bf y}(t)={\bf g}({\bf x}(t))=<\!{\bf y}|{\bf x}(t)\!>+{\bf \eta}(t)
\label{eq:ap2.9}
\end{equation}
where $< \,|{\bf x}>$ stands for the conditional average and
${\bf \eta}(t)$ is a noise term.
Inserting (\ref{eq:ap2.9}) into the first of (\ref{eq:ap2.3})
one obtains
\begin{equation}
\frac{{\rm d}{\bf x}}{{\rm d} t} = {\bf F}_{1}({\bf x})+
{\bf F}_{2}({\bf x},{\bf y})={\bf F}_{1}({\bf x})+
{\bf F}_{2}({\bf x},<\!{\bf y}|{\bf x}\!>) + \delta {\bf F}_{2}({\bf x})
\label{eq:ap2.10}
\end{equation}
where
\begin{equation}
\delta F_{2,i} = \left. \sum_{j,k} {\partial^2 F_{2,i} \over
\partial y_j \partial y_k} \right|_{y=<y|x>} \langle \eta_j \eta_k \rangle
\label{eq:ap2.11}
\end{equation}
In the Lorenz '96 model (\ref{eq:3.6}), because of the linear coupling
between the different scales, the terms $\delta {\bf F}_{2}$ are
absent and one has a close model for the large scale variables
\begin{equation}
\frac{{\rm d}{\bf x}}{{\rm d} t} = {\bf F_{1}}({\bf x})
+{\bf F_{2}}({\bf x},<\!{\bf y}|{\bf x}\!>)
\label{eq:ap2.12}
\end{equation}

The ansatz (\ref{eq:ap2.9}) is well verified in the numerical
simulations.  We have computed the $\lambda_{TM}(\delta)$ by using a
best fit for ${\bf F_{2}}$ and we have obtained a good reproduction of
the $\lambda_{TT}(\delta)$ for large $\delta$.  In the Lorenz '96
model (\ref{eq:3.6}), where the coupling between slow and fast
variables is practically linear, one has that $F_{2,k}({\bf x},
<\!{\bf y}|{\bf x}\!>)=\sum_{j=1,J} <\!y_{j,k}|x_k\!> \simeq \nu_e
x_k$.

Now we will discuss the case of the Shell Model parameterization
which pertains to the general issue of the subgrid-scale modelization.
The literature on this field and the related problems
(e.g. closure in fully developed turbulence) is enormous and
we do not pretend to discuss here in details this field.
Let us only recall the basic idea introduced over a century ago by
Boussinesq, and later developed further by Taylor, Prandtl and Heisenberg
-- to cite some of the most famous ones-- for fully developed turbulence
(Frisch 1995).
In a nutshell the idea is to mimic the energy flux from the large to the
small scales (in our terms from slow to fast variables) by an effective
dissipation: the effect of the small scales on the
large ones can be modeled as an  enhanced molecular viscosity.

By simple dimensional arguments one can argue that the effects of small
scales can be replaced by an effective viscosity at scales $r$, given by
\begin{equation}
\nu^{(e)} \sim r \delta v (r)
\label{eq:ap2.7}
\end{equation}
where $\delta v(r)$ is the velocity fluctuation on the scale $r$.

The above argument for the Shell Model (\ref{eq:4.1}) gives (Benzi et
al. 1998):
\begin{equation}
\nu^{(e)}_{n} = \kappa {|u_n| \over k_n}
\label{eq:ap2.8}
\end{equation}
where $\kappa \sim O(1)$ is an empirical constant.
 From eq. (\ref{eq:ap2.7}) one could naively think to use
dimensional argument {\it \'a la} Kolmogorov to set a constant eddy
viscosity $\nu^{(e)}_{n} \sim k_n^{-4/3}$. In this way one forgets the dynamics
and this can cause numerical blow up. More
sophisticated arguments that do not include the dynamics lead to similar
problems.

Let us remark that the parameterization (\ref{eq:ap2.8}) is not
exactly identical to those obtained by closure approaches where
the eddy viscosity is given in terms of averaged quantities. In
our case this would mean to write $\langle u_n^2 \rangle^{1/2}$
instead of $|u_n|$ in (\ref{eq:ap2.8}).

After this discussion it is easy to recognize that the
parameterization in terms of conditional averages introduced for the
Lorenz '96 model is, {\it a posteriori}, an eddy viscosity model.


\newpage

\centerline{FIGURE CAPTIONS}

\begin{figure}[hbt]
\epsfbox{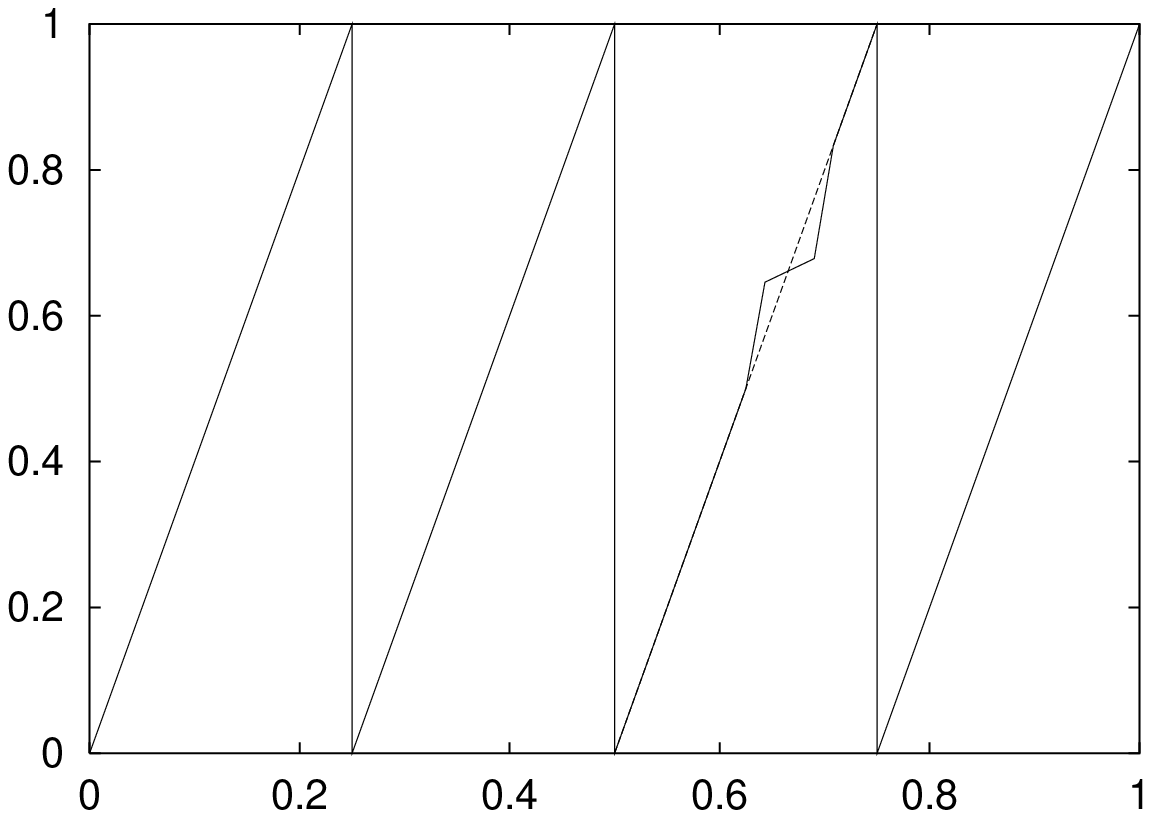}
\caption{
}
\label{fig0}
\end{figure}

\begin{figure}[hbt]
\epsfbox{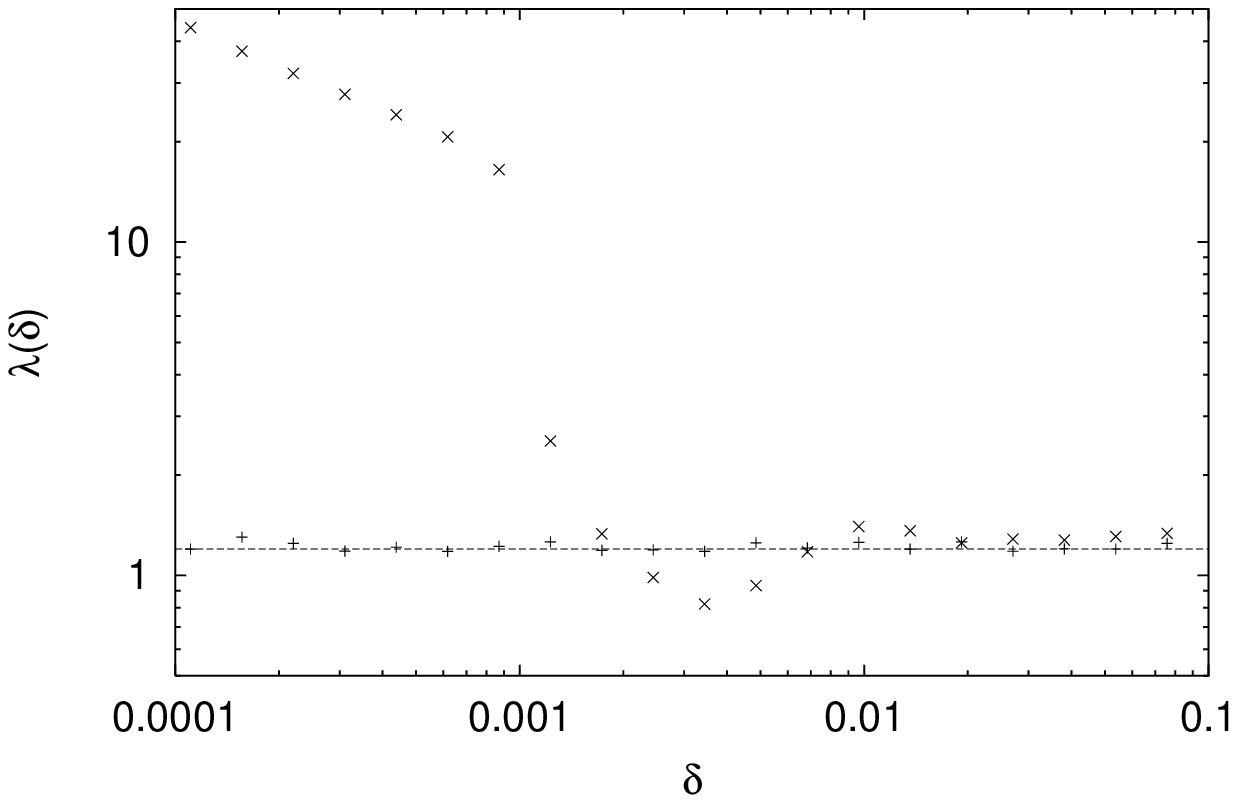}
\caption{
}
\label{fig1}
\end{figure}

\begin{figure}[hbt]
\epsfbox{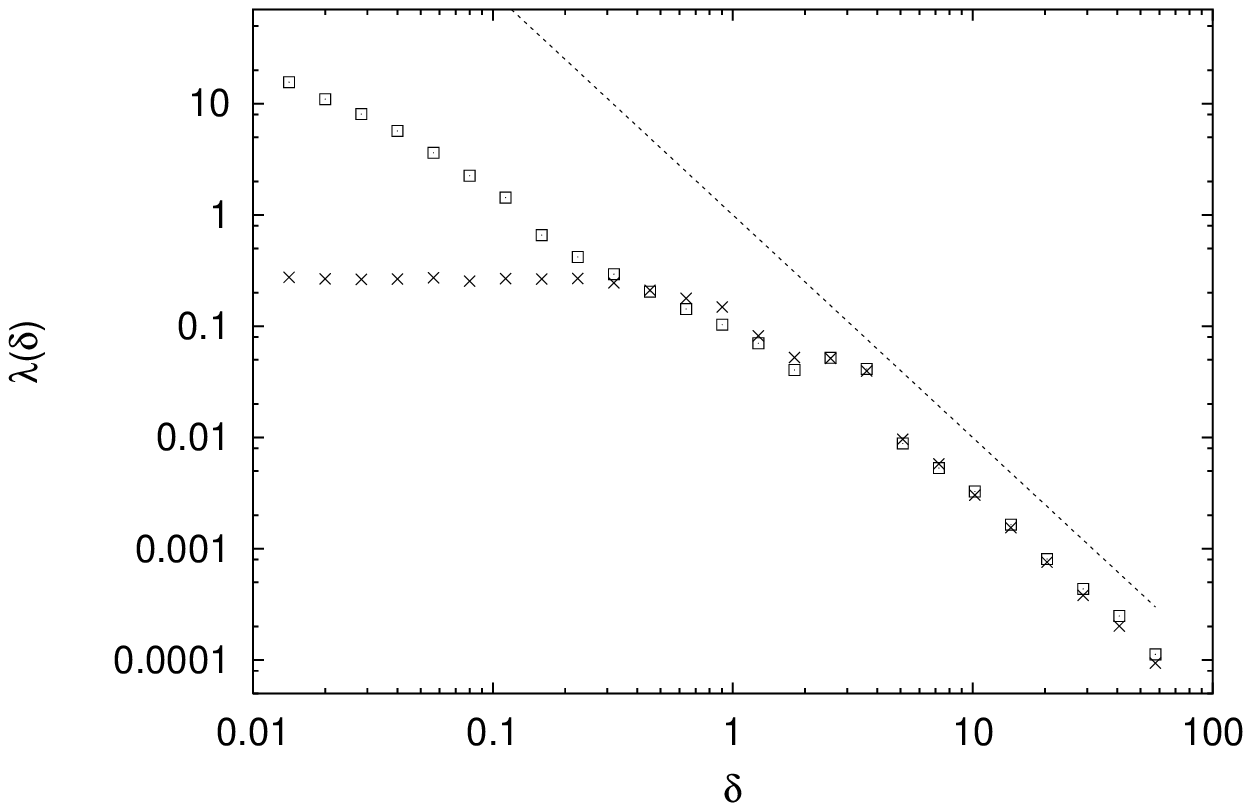}
\caption{
}
\label{fig2}
\end{figure}

\begin{figure}[hbt]
\epsfbox{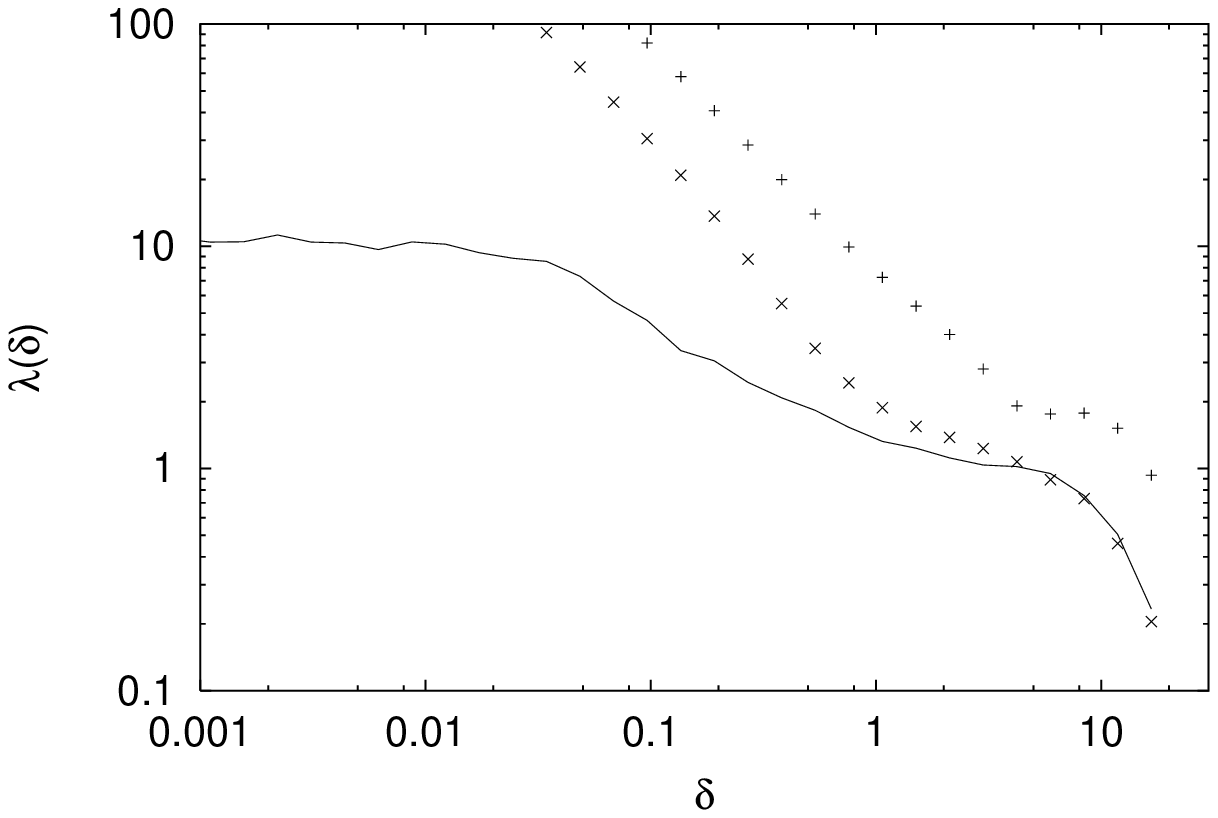}
\caption{
}
\label{fig3}
\end{figure}

\begin{figure}[hbt]
\epsfbox{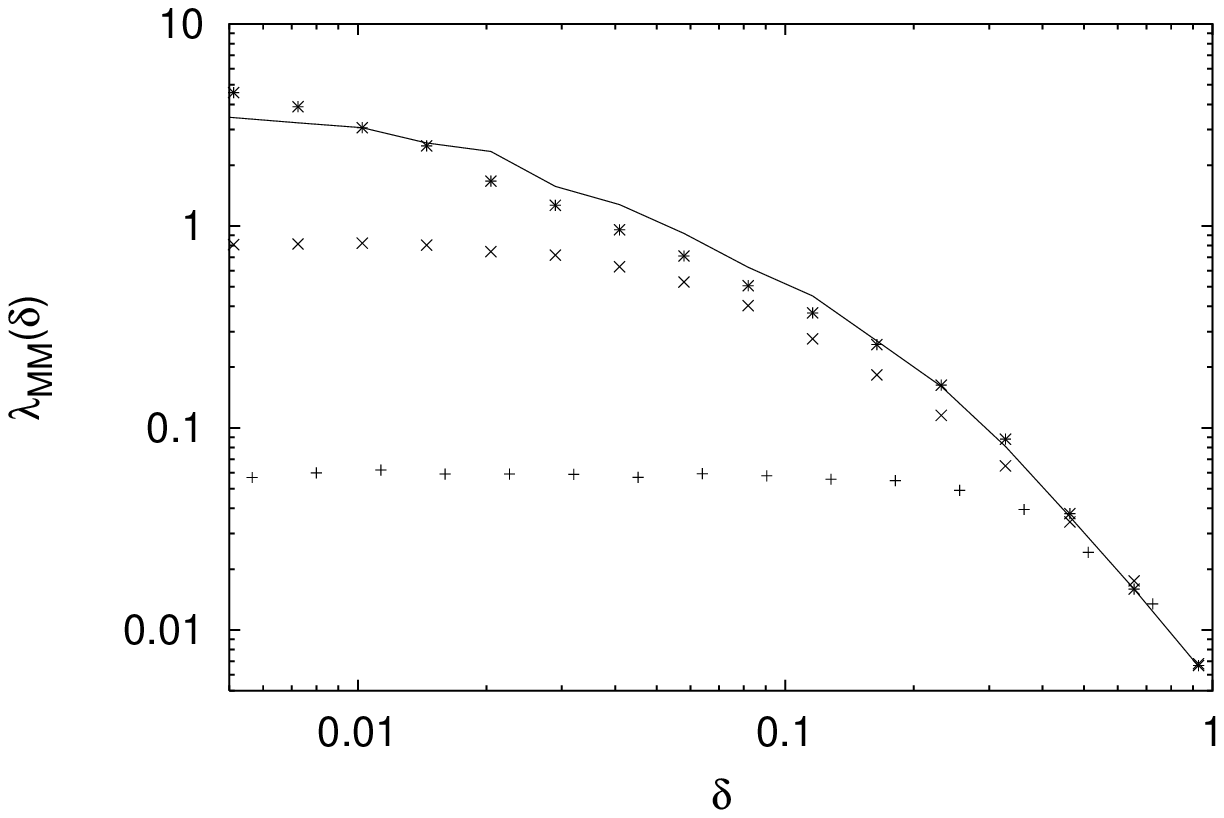}
\caption{
}
\label{fig4.1}
\end{figure}

\begin{figure}[hbt]
\epsfbox{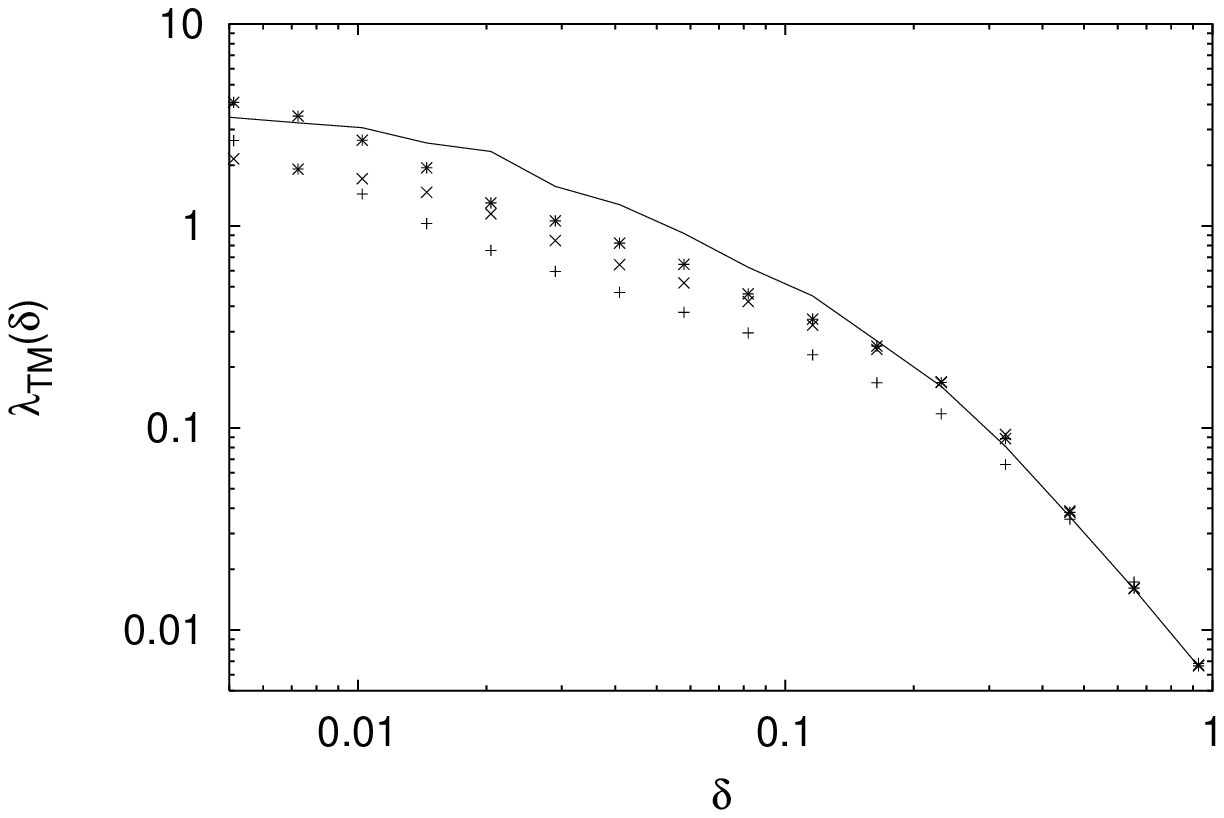}
\caption{
}
\label{fig4.2}
\end{figure}

\begin{itemize}

\item [FIGURE \ref{fig0}:] 
The map $f_p$ of equation (\ref{eq:ap0}) (solid line) and the
original chaotic map $f$ (dashed line).

\item [FIGURE \ref{fig1}:] 
Finite Size Lyapunov Exponents
$\lambda_{TT}(\delta)$ ($+$)
and $\lambda_{TM}(\delta)$ ($\times$) versus $\delta$ for
the Lorenz model (\ref{eq:2.1}) with $\sigma=c=10$,
$b=8/3$, $R=45$ and $\epsilon=0.001$. The dashed line represents the
leading Lyapunov exponent for the unperturbed system ($\lambda \approx 1.2$).
The statistics is over $10^4$ realizations.

\item [FIGURE \ref{fig2}:] 
$\lambda_{TT}(\delta)$ (crosses, $\times$) and
$\lambda_{TM}(\delta)$ (open squares, $\Box$) versus $\delta$
for the Rayleigh-B\'enard model (\ref{eq:2.6}) with
$C=0.5$, $k_L=1$, $\omega_L=1$, $B_L=0.3$,
$k_S=4$, $\omega_S=4$, $B_S=0.3$ and $\epsilon=0.125$.
The straight line indicates the $\delta^{-2}$ slope.
The statistics is over $10^4$ realizations.

\item [FIGURE \ref{fig3}:] Finite Size Lyapunov Exponents for the
Lorenz '96 model $\lambda_{TT}(\delta)$ (solid line) and
$\lambda_{TM}(\delta)$ versus $\delta$ obtained by dropping the fast
modes ($+$) and with eddy viscosity parameterization ($\times$) as
discussed in (\ref{eq:3.7}) and (\ref{eq:3.8}).  The parameters are
$F=10$, $K=36\,,\;J=10$, $\nu=1$ and $c=b=10$, implying that the
typical $y$ variable is $10$ times faster and smaller than the $x$
variable.  The value of the parameter $\nu_e=4$ is chosen after a
numerical integration of the complete equations as discussed in
Appendix \ref{app:2}. The statistics is over $10^4$ realizations.

\item [FIGURE \ref{fig4.1}:] 
The FSLE for the eddy-viscosity shell model
(\ref{eq:4.6}) $\lambda_{MM}(\delta)$ at
various resolutions $N_M=9(+),15(\times),20(\ast)$.
For comparison it is drawn the
FSLE $\lambda_{TT}(\delta)$ (continuous line). Here $\kappa=0.4$,
$k_0=0.05$.

\item [FIGURE \ref{fig4.2}:] 
The FSLE between the eddy-viscosity shell model 
and the full shell model $\lambda_{TM}(\delta)$, at various
resolutions $N_M=9(+),15(\times),20(\ast)$. For comparison it is drawn the
FSLE $\lambda_{TT}(\delta)$ (continuous line).
The total number of shell for the complete model is $N=24$, with
$k_0=0.05$, $\nu=10^{-7}$.

\end{itemize}

\newpage 

\end{document}